\documentclass[10pt, sigconf, letterpaper, nonacm]{acmart}

\usepackage{breakurl}
\usepackage{subfigure}

\usepackage{float}
\usepackage{url}
\usepackage{xcolor}
\usepackage{balance}
\usepackage{xspace}
\usepackage{amsmath}
\usepackage{hyperref}
\usepackage{amsfonts}%
\usepackage{multirow}
\usepackage{graphicx}
\usepackage{booktabs}
\usepackage{caption}
\usepackage{array}
\usepackage{enumitem}
\setlist[itemize]{left=0pt}
\setlist[enumerate]{left=0pt}

\usepackage{color, colortbl}
\usepackage[first=0,last=9]{lcg}

\definecolor{Gray}{gray}{0.9}
\definecolor{LightCyan}{rgb}{0.88,1,1}

\hypersetup{
    colorlinks=true,
    linkcolor=blue,
    filecolor=magenta,
    urlcolor=cyan,
    pdfpagemode=FullScreen,
}

\usepackage{acronym}

\acrodef{ZKP}[ZKP]{Zero-knowledge proof}

\acrodef{DeFi}[DeFi]{Decentralized Finance}

\acrodef{CeFi}[CeFi]{Centralized Finance}

\acrodef{PoS}[PoS]{Proof-of-Stake}

\acrodef{PoW}[PoW]{Proof-of-Work}

\acrodef{ENS}[ENS]{Ethereum Name Service}

\acrodef{TC}[TC]{Tornado.Cash}

\acrodef{TN}[TN]{Typhoon.Network}

\acrodef{TP}[TP]{Typhoon.Cash}

\acrodef{CEX}[CEX]{Centralized Exchange}

\acrodef{DEX}[DEX]{Decentralized Exchange}

\acrodef{P2P}[P2P]{peer-to-peer}

\acrodef{DApp}[DApp]{Decentralized Application}

\acrodef{MEV}[MEV]{Miner Extractable Value}

\acrodef{BEV}[BEV]{Blockchain Extractable Value}

\acrodef{FaaS}[FaaS]{Front-running as a Service}

\acrodef{OFAC}[OFAC]{Office of Foreign Assets Control}

\acrodef{AM}[AM]{anonymity mining}

\acrodef{BSC}[BSC]{Binance Smart Chain}

\acrodef{ETH}[ETH]{Ethereum}

\acrodef{CLI}[CLI]{command line interface}

\begin{document}
\title{Time to Bribe: Measuring Block Construction Markets}

\author{
Anton Wahrstätter\textsuperscript{1}, 
Liyi Zhou\textsuperscript{2}\textsuperscript{4}, 
Kaihua Qin\textsuperscript{2}\textsuperscript{4}, \\ 
Davor Svetinovic\textsuperscript{1}\textsuperscript{3}, 
Arthur Gervais\textsuperscript{4}\textsuperscript{5}}

\affiliation{
	\textsuperscript{1}Vienna University of Economics and
 Business
    \country{}
	\textsuperscript{2}Imperial College London
    \country{}
    \textsuperscript{3}Khalifa University
    \country{}
	\\
	\textsuperscript{4}Berkeley Center for Responsible, Decentralized Intelligence (RDI)
    \country{}
	\textsuperscript{5}University College London
}
\renewcommand{\shortauthors}{Wahrstätter et al.}


\begin{abstract}
With the emergence of Miner Extractable Value (MEV), block construction markets on blockchains have evolved into a competitive arena. Following Ethereum's transition from Proof of Work (PoW) to Proof of Stake (PoS), the Proposer Builder Separation (PBS) mechanism has emerged as the dominant force in the Ethereum block construction market.

This paper presents an in-depth longitudinal study of the Ethereum block construction market, spanning from the introduction of PoS and PBS in September 2022 to May 2023. We analyze the market shares of builders and relays, their temporal changes, and the financial dynamics within the PBS system, including payments among builders and block proposers~---~commonly referred to as \emph{bribes}. We introduce an MEV-time law quantifying the expected MEV revenue wrt.\ the time elapsed since the last proposed block. We provide empirical evidence that moments of crisis (e.g.\ the FTX collapse, USDC stablecoin de-peg) coincide with significant spikes in MEV payments compared to the baseline.

Despite the intention of the PBS architecture to enhance decentralization by separating actor roles, it remains unclear whether its design is optimal. Implicit trust assumptions and conflicts of interest may benefit particular parties and foster the need for vertical integration. MEV-Boost was explicitly designed to foster decentralization, causing the side effect of enabling risk-free sandwich extraction from unsuspecting users, potentially raising concerns for regulators.
\end{abstract}


\maketitle

\section{Introduction}\label{sec:introduction}
Over the past decade, decentralized ledgers like Bitcoin and Ethereum have mainly depended on peer-to-peer networks to disseminate transaction data. This layer responsible for information distribution has seen significant changes recently. At first, miners started offering ports to allow traders to submit their transaction data directly. Afterward, companies such as Flashbots implemented centralized intermediaries to manage sealed-bid auctions for transaction inclusion. Collaborating with miners by shifting the complexity of the auction process away from them, Flashbots was able to increase its market footprint swiftly.

Ethereum's evolution from a Proof of Work to a Proof of Stake consensus mechanism involved a shift towards a more complex Proposer Builder Separation system. Essentially, the PBS mechanism differentiates between various participants while fundamentally reducing the power of PoS proposers in determining which transactions are included within their proposed blocks. The PBS system comprises three unique roles: searchers, builders, and relays, all with distinct responsibilities. These specialized agents work together to profit from fleeting arbitrage opportunities, commonly known as Miner Extractable Value. This value primarily stems from pricing inconsistencies across Decentralized Finance (DeFi) platforms~\cite{qin2021cefi}, discrepancies between the pricing of decentralized and centralized exchanges, or price variances across different blockchain platforms~\cite{qin2022quantifying}.

In the PBS framework, the primary role of searchers is to identify MEV opportunities and then assemble transaction bundles that exploit these prospects. Once these bundles are assembled, they are handed over to builders, who commence crafting blocks that optimize revenue generation. In parallel, relays scrutinize these blocks, pinpoint the most profitable ones, and relay those to the proposers. The block proposer chosen to propose the upcoming block can utilize the MEV-Boost software to receive externally built blocks to fulfill its role effectively~\cite{flashbots2022introduction}. Proposers, formerly referred to as miners, are faced with a decision: they must choose whether to accept blocks suggested by a relay or to propose blocks they've constructed locally, incorporating transactions either from Ethereum's openly accessible peer-to-peer layer or transmitted directly to them through other parties. In this context, transactions pending verification typically reside within a data structure known as the mempool.

In this paper, we conduct a longitudinal study of the block construction market within Ethereum. Our observations span from the 1st of September, 2022, to the 1st of May, 2023. Ethereum notably modified its consensus mechanism and implemented Proof of Stake and Proposer Builder Separation on September 15th. We note that MEV-Boost's market share rose to 90\% within roughly two months. Interestingly, the market for block builders seemed to maintain a certain level of diversity throughout our measurement period, with no single builder consistently commanding more than 20\% of the market share. As for relays, the Flashbots relay reached over 60\% market share in November 2022, but this figure fell to around 25\% by May 1st. Due to the anonymity of many builders, comparable to mining pools, it remains uncertain which entities own which builder or relay.

This paper makes the following contributions:

\begin{description}
    \item[Measuring MEV-Boost \& Money Flows] We carry out the first longitudinal measurement of the PBS market on Ethereum, from September 2022 to May 2023. We dissect the market shares of builders and relays over time, and present a visual representation of the network topology. We found that the total amount of ETH transferred to block proposers via MEV-Boost was 160k ETH (320M USD). We also delve into the monetary flows within PBS, such as payments between builders, proposers, and block proposers.
    
    Our findings show that only 1\% of the MEV transactions exceed 1.36 ETH, and that 20\% of the highest MEV payments account for 72\% of the total revenue. These observations exacerbate considerations related to the security of the consensus mechanism.
    
    \item[MEV-Time Law] We introduce the MEV-Time Law, which measures the anticipated financial MEV earnings relative to the time elapsed within an Ethereum slot. We identify a peak MEV value at second 2.78 of a slot age, where, on average the MEV bid is 107\% higher than 2 seconds before the commencement of a slot.
    
    \item[Timing Attacks] We uncover quantifiable evidence of temporal variations among proposers, suggesting that certain proposers may delay processing to receive more valuable MEV payments compared to others. Bitcoin Suisse seems to set a standard in terms of consistency, while Kraken or Binance appears more flexible regarding the timing within a slot. 
    
    \item[Crisis increases MEV] Our study presents empirical evidence that moments of crisis amplify MEV revenue. Specifically, events like the FTX collapse and the de-pegging of the USDC stablecoin led to respective increases in MEV revenues by 400\% and 1000\% for several days, when compared to the baseline.
\end{description}

\section{Background}\label{sec:background}

\subsection{Ethereum}
Ethereum is a blockchain platform that enables users to carry out transactions and use smart contracts without requiring a centralized authority. This platform is built upon distributed ledger technology and is maintained by a decentralized network of nodes. Each node holds a copy of the Ethereum blockchain, a distributed record of all transactions within the network. Transactions are disseminated to the peer-to-peer layer of the network, where every node receives and validates them. Once a transaction is validated, it is appended to the blockchain and becomes a permanent part of the ledger's historical record. Previously, Ethereum utilized a consensus mechanism known as Proof of Work (PoW), which required miners to compete in solving mathematical problems to add new blocks to the blockchain~\cite{bano2019sok}. However, as of September 2022, Ethereum has shifted to Proof of Stake, where the consensus voting power is determined by a participant's stake in the system~\cite{schwarz2022three,neu2022two}. Ether, the native cryptocurrency of Ethereum, is used for various purposes, including the payment of transaction fees. Smart contracts on Ethereum are self-executing programs that allow all participants to verify the execution of the computer program, enabling the implementation of agreements without the need for centralized trust assumptions. These smart contracts are executed by the Ethereum Virtual Machine (EVM)~\cite{wood2014ethereum}. Developers use Ethereum to create decentralized applications (DApps), including ERC-20 tokens, which can operate independently without intermediaries. Tokens represent a virtual asset symbolizing any scarce item or currency, such as votes in a decentralized system or collectible items in a metaverse game. Decentralized Exchanges (DEXs) are a significant practical application of Ethereum's smart contracts. DEXs facilitate peer-to-peer token trading based on simple rules outlined in the smart contract, eliminating the need for a central authority. One of the most popular DEX design models are Automated Market Makers (AMMs), which use a mathematical formula to determine the price for a pair of tokens on-chain.

\subsection{Miner Extractable Value}
MEV refers to the potential profit that can be garnered from transactions included within an Ethereum block. MEV extraction entails the strategic rearrangement, inclusion, or exclusion of transactions during block creation to accrue extra profit beyond the conventional block rewards.

In Ethereum, a unit known as ``gas'' calculates the computational cost of executing a transaction on the network. A Priority Gas Auction (PGA) is a market-based mechanism that determines the computation price for network transactions. When a user initiates a transaction, they specify the maximum price they are willing to pay for its inclusion into the blockchain. The PGA then matches this transaction with the lowest gas price offered by a miner, encouraging users to set competitive gas prices to expedite their transaction processing, thus potentially enhancing miners' transaction fee earnings. Transaction sequencing plays a pivotal role in extracting MEV and primarily involves two methodologies: front-running and back-running. Front-running occurs when an MEV-seeking entity purposely pays a higher gas price than a specific transaction, thus ensuring their transaction is included in the blockchain before the targeted one. Conversely, back-running involves an MEV-seeking party positioning their transaction immediately after the target's transaction to gain from it. Such transaction ordering not only enables these two forms of MEV extraction, but it also facilitates more intricate strategies like sandwich attacks and cyclic arbitrage. In a sandwich attack, the MEV extractor places transactions both before and after the targeted transaction to seize value. Cyclic arbitrage, on the other hand, leverages price discrepancies across different Decentralized Exchanges (DEXs) by executing a series of swaps. MEV extraction often leads to heightened competition among extractors for the same opportunities, causing elevated gas fees, network congestion, and extensive block space usage. Furthermore, MEV poses a threat to the inherent consensus security of Ethereum, a risk that has been quantified in related work~\cite{zhou2021just}. 

\paragraph{Flashbots}
Flashbots is a for-profit company that furnishes a server for conducting gas price auctions. Specifically, Flashbots' auctions facilitate a communication channel between Ethereum users and block proposers, enabling users to submit their transactions to an orthogonal mempool to the P2P mempool. Flashbots promises informally, but does not commit contractually, to keep the flashbots mempool and auction data private from third parties. Technically, this auction setup initially started as a partial modification to the go-ethereum client (referred to as mev-geth), and has since evolved to align with Ethereum's Proof of Stake consensus protocol.

The auction uses a transaction pool combined with a sealed-bid auction mechanism for block space allocation. It functions as a first-price sealed-bid auction, allowing those seeking to extract Miner Extractable Value to convey their bids and transaction order preferences without paying for unsuccessful bids and without passing through the P2P mempool. This auction mechanism is intended to optimize proposer payouts, while non-winning participants may retain their order privacy due to the sealed auction format. While it's not contractually obligatory, the auction claims to offer assurances such as pre-trade privacy and failed-trade privacy. Payments, often referred to as ``bribes,'' can be made directly to the proposers or indirectly via increased gas fees.

\section{MEV-Boost}
\begin{figure}[htb]
\begin{center}
\includegraphics[width=\columnwidth]{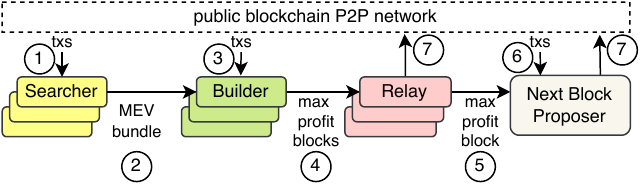}
\caption{Schematic representation of transaction and block propagation with Proposer Builder Separation. \textit{(1)} Searchers receive transactions from the P2P layer and generate transaction bundles using their specific MEV extraction knowledge. \textit{(2)} These bundles are then sent to one or more builders. \textit{(3)} Builders, who also receive transactions from the P2P layer, bundle blocks considering the transactions and bundles from searchers, guided by their local profit maximization algorithm. \textit{(4)} Builders connect with relays and send new maximum profit blocks to these relays as they're discovered. \textit{(5)} Upon request, relays share the status of the maximum profit bid with the next block proposer. \textit{(6)} The block proposer, who receives transactions from the P2P layer as well, decides which block to mine based on the relay information and their own interests. \textit{(7)} If the block proposer chooses the block from the relay, they return the signed block header, prompting the relay to share the actual block}
\label{fig:mev-boost-architecture}
\end{center}
\end{figure}

MEV-Boost represents a particular implementation of PBS, serving as a practical embodiment of the PBS framework. The initial MEV-Boost software was developed and released under an open-source license by Flashbots. Since its introduction, it has been adopted by numerous participants in the block construction market. MEV-Boost operates as an opt-in mechanism for block proposers, allowing them to access more profitable blocks meticulously constructed by proficient entities. A broad architectural illustration of the MEV-Boost system is provided in Figure~\ref{fig:mev-boost-architecture}.

\subsection{System Model}
PBS distributes the responsibilities of transaction bundling, sequencing, and block creation among the following parties.

\begin{description}
\item[searchers] These entities, driven by profit, crawl the blockchain state and other data sources to find lucrative opportunities such as arbitrage~\cite{mclaughlinlarge}, liquidations~\cite{qin2021empirical,qin2023mitigating}, and sandwich attacks~\cite{zhou2021high,wang2022impact,heimbach2022eliminating}.
\item[builders] The primary role of builders is to retrieve bundles of transactions from searchers and create blocks. Block proposers, in turn, select the most valuable blocks the builders offer for further propagation.
\item[relays] Relays act as intermediaries between the block proposer and the block builder. Their main function is to transmit the most profitable block to the proposer while establishing the necessary trust between the two parties.
\item[proposers] Proposers are entities that stake their ETH to propose and vouch for blocks. In the context of the former Proof of Work paradigm, proposers can be likened to miners. In this paper, we primarily use the term ``proposers'' and largely ignore the process of block attestation.
\end{description}




\subsection{Data Collection}

\begin{figure}[htb]
\begin{center}
\includegraphics[width=0.95\columnwidth]{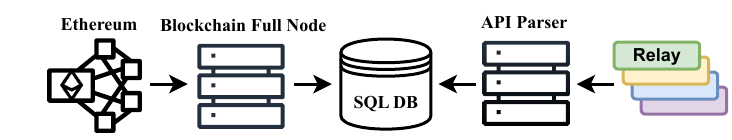}
\caption{Data Collection pipeline joining blockchain data (execution \& consensus layer) with data retrieved from the ``builder\_blocks\_received'' and ``proposer\_payload\_delivered'' endpoints of each operating MEV-Boost relay.\vspace{-0.3cm}}
\label{fig:data-collection}
\end{center}
\end{figure}

To examine the Ethereum ecosystem, we collect data about various players such as block proposers, block relayers, block builders, and the blockchain itself. This information is gathered using Ethereum execution and consensus layer nodes (as shown in Figure~\ref{fig:data-collection}). This detailed information is vital in helping us understand how these different participants interact and their impact on the consensus layer.

In our research, we employ Prysm and Geth nodes to gather data related to the execution and consensus layer of Ethereum. This approach enables us to collect comprehensive information about Proof-of-Stake (PoS) slots and their associated block proposers. When considering the roles of external entities, such as block builders and relay operators, we utilize the Relay Data API, which each active relay provider implements. Specifically, we make use of the "builder\_blocks\_received" and "proposer\_payload\_delivered" endpoints, as provided in the \href{https://flashbots.github.io/relay-specs}{Flashbots Relay Specifications}, to acquire data about the parties participating in the external block building market. Our primary focus is on the blocks that block builders deliver to proposers.

To comprehensively explore the ecosystem, we connect to every existing relay provider as of Mai 2023, including Flashbots, Ultra Sound, Agnostic, BloXroute, Blocknative, and others. Our final dataset contains $1,624,885$ blocks, including every block from the launch of PBS until 23:59:59 UTC on the 1st of May 2023. Our analysis reveals that external block builders constructed $1,312,852$ blocks ($80,8\%$). Conversely, their respective proposers built $312,033$ blocks ($19,2\%$) locally. To guarantee reproducibility, we make our raw dataset available under an Open-Source license at \url{https://mevboost.pics/data.html}. To further enrich our dataset, we employ the \href{https://etherscan.io/labelcloud}{Label Word Cloud} to map Ethereum addresses to identified entities, thus providing a more comprehensive picture of the Ethereum ecosystem and enabling deeper analysis of participant roles and relationships.

Creating a comprehensive and reliable dataset to analyze Ethereum's MEV-Boost ecosystem indeed presents several challenges, most notably:

\paragraph{Parsing and extracting blockchain data:}
Although blockchain data is publicly accessible, parsing this data and extracting the required information is a complex task. Specialized tools and techniques are required to effectively handle this data, with careful attention to all the details. The increasing complexity of the Ethereum blockchain, particularly in its need to distinguish between the execution and consensus layers---each presenting unique challenges---calls for the separate parsing of each layer before interconnecting them. The endeavor becomes even more complex within the Ethereum PBS ecosystem. This system is rife with nuances and intricate details. Navigating these requires both meticulous precision and an expansive comprehension of the ecosystem. Recognizing these challenges is fundamental in shaping our data collection and analysis methods. This ensures that our approach remains robust and encompassing in nature, facilitating a comprehensive understanding of the ecosystem.

\paragraph{Including a complete set of relay providers:}

While it might seem enticing to concentrate on the more prominent entities in our analysis, it is vital to consider the inclusion of smaller relay providers as well. Overlooking them might risk painting a skewed picture of the overall distribution of the MEV-Boost software in the ecosystem. 
In the Ethereum-based proposer builder separation ecosystem, the role of smaller relay providers cannot be understated. Although these players may appear insignificant individually, they collectively constitute a significant portion of the network. Hence, their contribution to using the MEV-Boost software should not be disregarded.
Additionally, due to the vertical integration in this system, excluding smaller relays from the analysis would inevitably mean neglecting those builders that exclusively submit blocks to these particular relays.
Failure to incorporate them could lead to a distorted view of the MEV-Boost software's distribution and use, ultimately affecting our understanding of the overall Ethereum ecosystem.
Therefore, our methods must be tailored so that we are equally considerate of all entities, irrespective of their size, within the Ethereum PBS ecosystem.

\paragraph{Reliability of MEV-Boost data: }
Specific details about the MEV-Boost ecosystem, including bid data, are challenging to verify with certainty. This situation mandates a certain level of reliance on the accuracy of the data provided by individual relay API endpoints.
Given the inherent complexities of the MEV-Boost ecosystem, validating specific data such as bids becomes increasingly challenging. Unfortunately, a lack of reliable methods for verification exists, requiring researchers and analysts to place a significant amount of trust in the data obtained from individual relay API endpoints.
The necessity of this trust highlights an area of vulnerability in our data collection and analysis process. Therefore, we must recognize this limitation as we interpret the data.

\paragraph{Duplication of blocks:} Some blocks are reported to be delivered by multiple relays. This is expected because a builder may submit the same blocks to multiple relays, increasing the chance of accepting the block. However, this results in an inflated count of relayed blocks. To address this, the analysis in this paper counts blocks reported by multiple relays as having originated from each relay that claims to have delivered the respective block. Consequently, the total count of blocks per relay is reported to be larger than the actual sum of MEV-Boost blocks. However, this doesn't affect the reported market shares of the individual entities over time or figures concerning the monetary rewards.

\subsection{MEV-Boost Insights}
In the following, we will investigate our initial insights into the MEV-Boost ecosystem and the dynamics that PBS entails.

\begin{figure}[htb]
\begin{center}
\includegraphics[width=0.95\columnwidth]{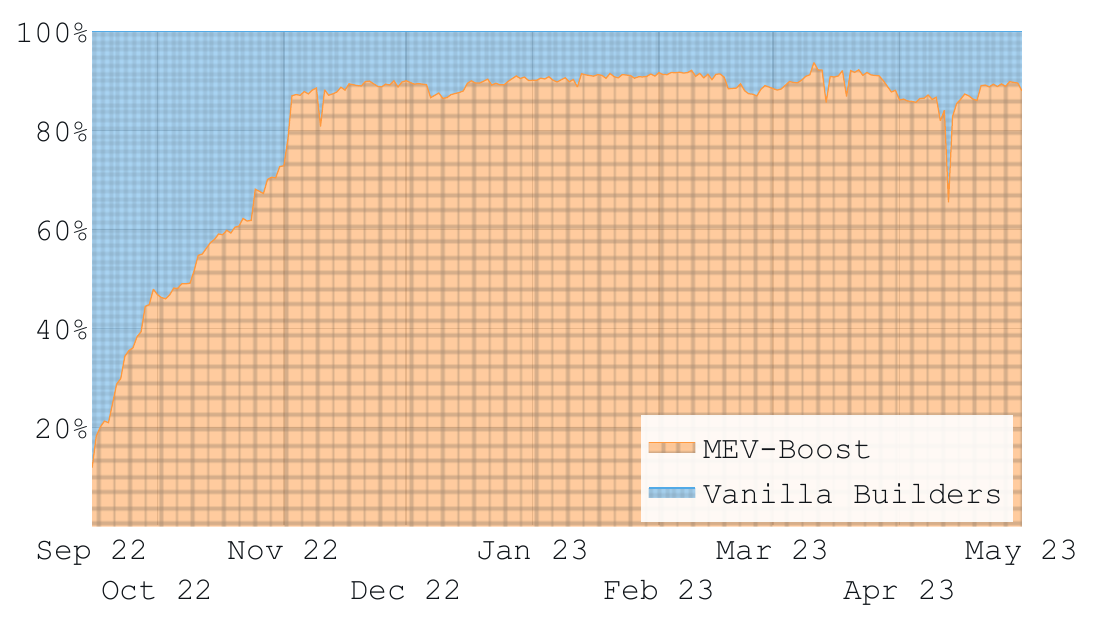}
\caption{MEV-Boost market share. We observe that MEV-Boost is a clearly dominant PBS implementation.}
\label{fig:mev-boost-market-share}
\end{center}
\end{figure}

\paragraph{Market Share}
Figure~\ref{fig:mev-boost-market-share} illustrates the significant and enduring market presence of MEV-Boost since the begin of our measurements. It appears that approximately 90\% of block proposers have chosen to employ MEV-Boost for their block building activities, attesting to its widespread acceptance and influence in the Ethereum ecosystem. Interestingly, the remaining 10\% of the block proposers have opted to forego the use of MEV-Boost. These participants, referred to as ``vanilla builders,'' continue to perform block building independently without leveraging the MEV-Boost software. This suggests that while MEV-Boost has achieved considerable adoption, a segment of the market still prefers traditional block-building.

\paragraph{MEV-Boost Payments}

\begin{figure}[htb]
\begin{center}
\includegraphics[width=0.95\columnwidth]{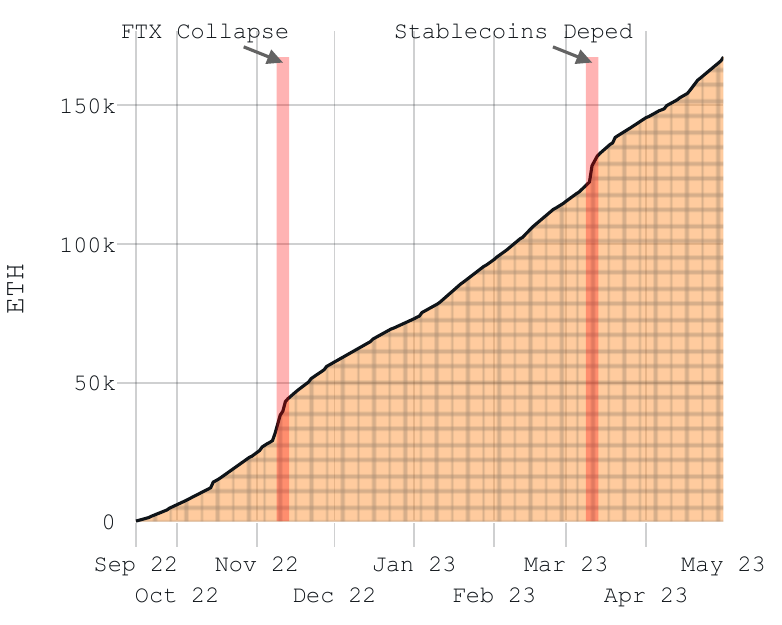}
\caption{Cumulated amount of ETH transferred to block proposers through MEV-Boost.\vspace{-0.3cm}}
\label{fig:mev-boost-cumulated-ETH}
\end{center}
\end{figure}

During our eight-month measurement period, we calculated that the total amount of ETH transferred to block proposers via MEV-Boost reached 160,000 ETH. Given the exchange rates at the time of writing, this amount equates to approximately 320 million USD, as shown in Figure~\ref{fig:mev-boost-cumulated-ETH}. Notably, this period witnessed two significant events in the blockchain and crypto space. The first was the collapse of the cryptocurrency exchange FTX. The second event was the de-pegging of the USDC stablecoin, a digital currency whose value is typically pegged to the US dollar.

\begin{figure}[htb]
\begin{center}
\includegraphics[width=0.95\columnwidth]{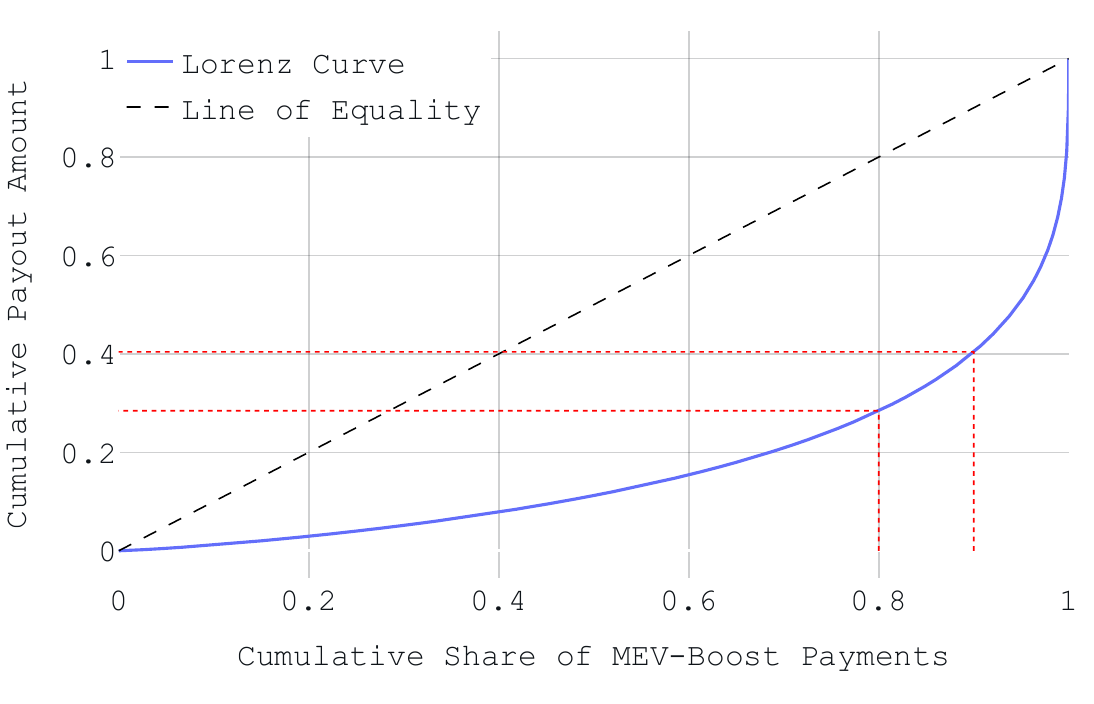}
\caption{MEV-Boost Payment Lorenz Curve, showing that 80\% of the payments have 28\% of the value, 90\% account for 39\% of the total MEV share.}
\label{fig:mev-boost-lorenz-curve}
\end{center}
\end{figure}

This part explores how payments are spread out in the MEV-Boost ecosystem. We do so by comparing the cumulative payout amount with the cumulative share of MEV-Boost payments, as shown in the Lorenz curve in Figure~\ref{fig:mev-boost-lorenz-curve}. The data indicates that a considerable 80\% of the lower MEV payments only make up 28\% of the overall value, and 90\% of the payments only account for 39\% of the total MEV value. This suggests a significant imbalance where a small fraction, just 20\%, of MEV payments capture a large portion, 72\%, of the revenue. Our data point to a significant discrepancy in the payout amounts of MEV transactions. This implies that when proposers propose a block, they essentially participate in a sort of lottery. More importantly, this stark imbalance in payment distribution disrupts the usual incentive schemes of blockchain networks, where transaction fee payments are generally expected to be more balanced. Additionally, other studies have shown that an overly high MEV, particularly when it surpasses the average block reward, can potentially compromise the security of the consensus mechanism~\cite{zhou2021just}.

\begin{figure}[htb]
\begin{center}
\includegraphics[width=0.95\columnwidth]{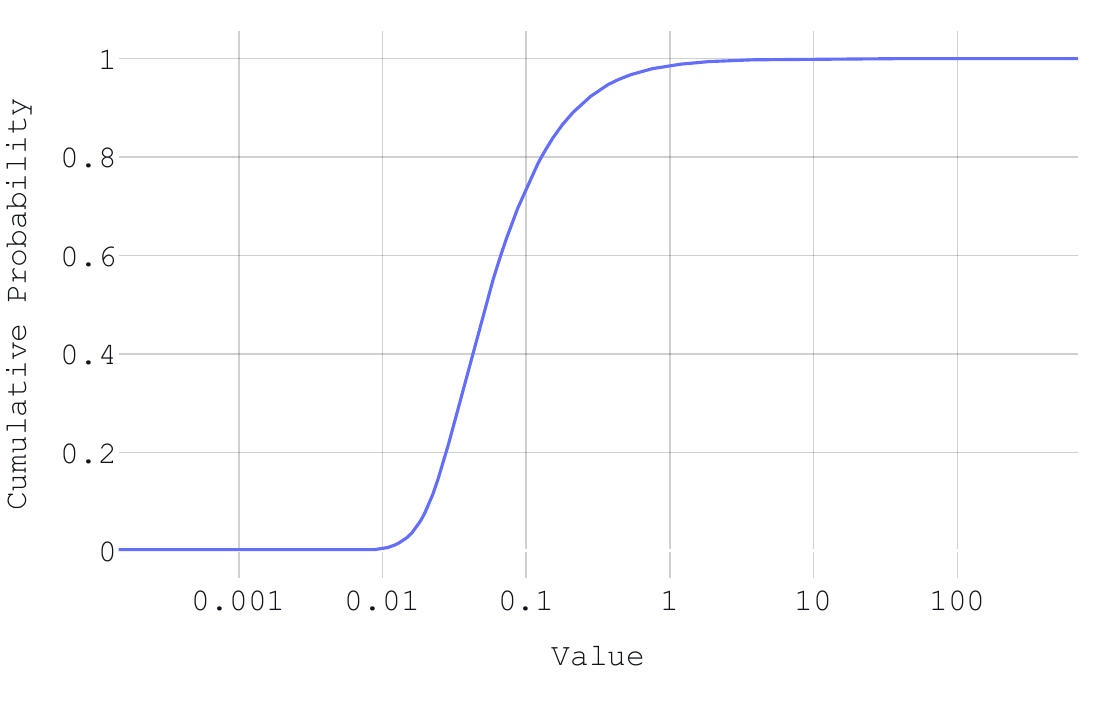}
\caption{MEV-Boost Payment CDF against their value in ETH. We find that only 1\% of the MEV transactions are above 1.36 ETH.}
\label{fig:mev-boost-payment-cdf}
\end{center}
\end{figure}

In a more in-depth investigation of the imbalance in MEV transaction values, we discover that a mere 1\% of MEV-Boost payments exceed 1.36 ETH, as depicted in Figure~\ref{fig:mev-boost-payment-cdf}. This finding further emphasizes the payment disparities in MEV, leading to a competitive environment where searchers, builders, relayers, and proposers are motivated to compete for the limited number of high-value blocks.

\paragraph{FTX collapse}
In November 2022, the international crypto community was affected by the swift downfall of the well-known cryptocurrency exchange FTX, established by Sam Bankman-Fried. The crisis was set off by media reports cautioning about leverage and solvency issues linked to Alameda Research, an entity affiliated with FTX. Within a few days, FTX was facing a liquidity crisis and made a failed attempt to secure a bailout from Binance. Consequently, Bankman-Fried, the CEO, resigned and FTX filed for bankruptcy. An alleged hacking incident occurred shortly after, resulting in the theft of tokens worth hundreds of millions. 

\begin{figure}[htb]
\begin{center}
\includegraphics[width=0.95\columnwidth]{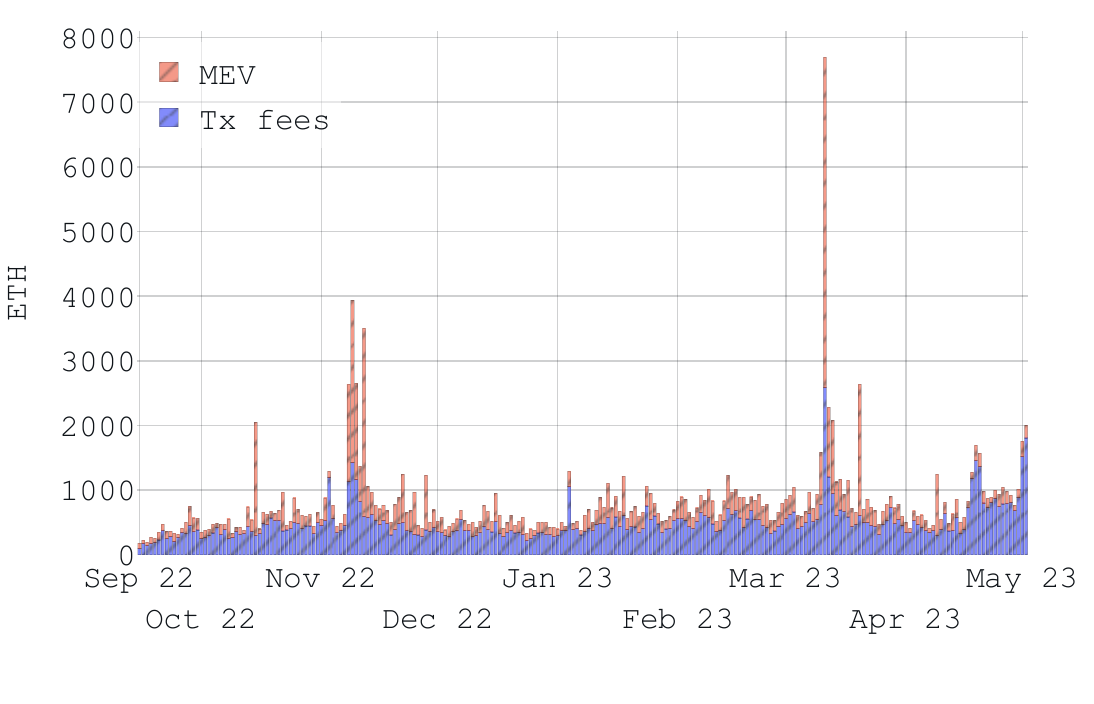}
\caption{MEV-Boost and transaction fee revenue per day. We can again identify how the MEV rewards and transaction fees rise by nearly an order of magnitude during market shocks (i.e., the FTX collapse and the USDC de-peg).\vspace{-0.3cm}}
\label{fig:mev-boost-fees-vs-mev}
\end{center}
\end{figure}

\paragraph{USDC de-peg}
In March 2023, the collapse of Silicon Valley Bank (SVB) has led to turmoil in the stablecoin market, as Circle's USDC stablecoin lost its peg to the U.S.\ dollar following a wave of investor withdrawals. Circle, the founder of USDC, disclosed that it had \$3.3 billion invested in SVB, causing panic among investors and leading to a significant drop in USDC's market cap. Other major stablecoins, such as DAI, also de-pegged to 0.9 USD as a reaction to this event. USDC swiftly regained its peg when the traditional markets re-opened after the weekend. To shed further light on the distinction between standard blockchain transaction fees and MEV-Boost payments, focusing on MEV revenue, we present Figure~\ref{fig:mev-boost-fees-vs-mev}. It is evident from this that MEV is becoming a significant portion of the proposers' income. In addition, while transaction fees may vary by a factor of 2 to 4, MEV payments can increase by more than tenfold, compared to the average. This signifies that the efforts of MEV-Boost to professionalize MEV extraction and its auction mechanism intensify the uneven distribution of MEV rewards.

\begin{figure}[htb]
\begin{center}
\includegraphics[width=0.95\columnwidth]{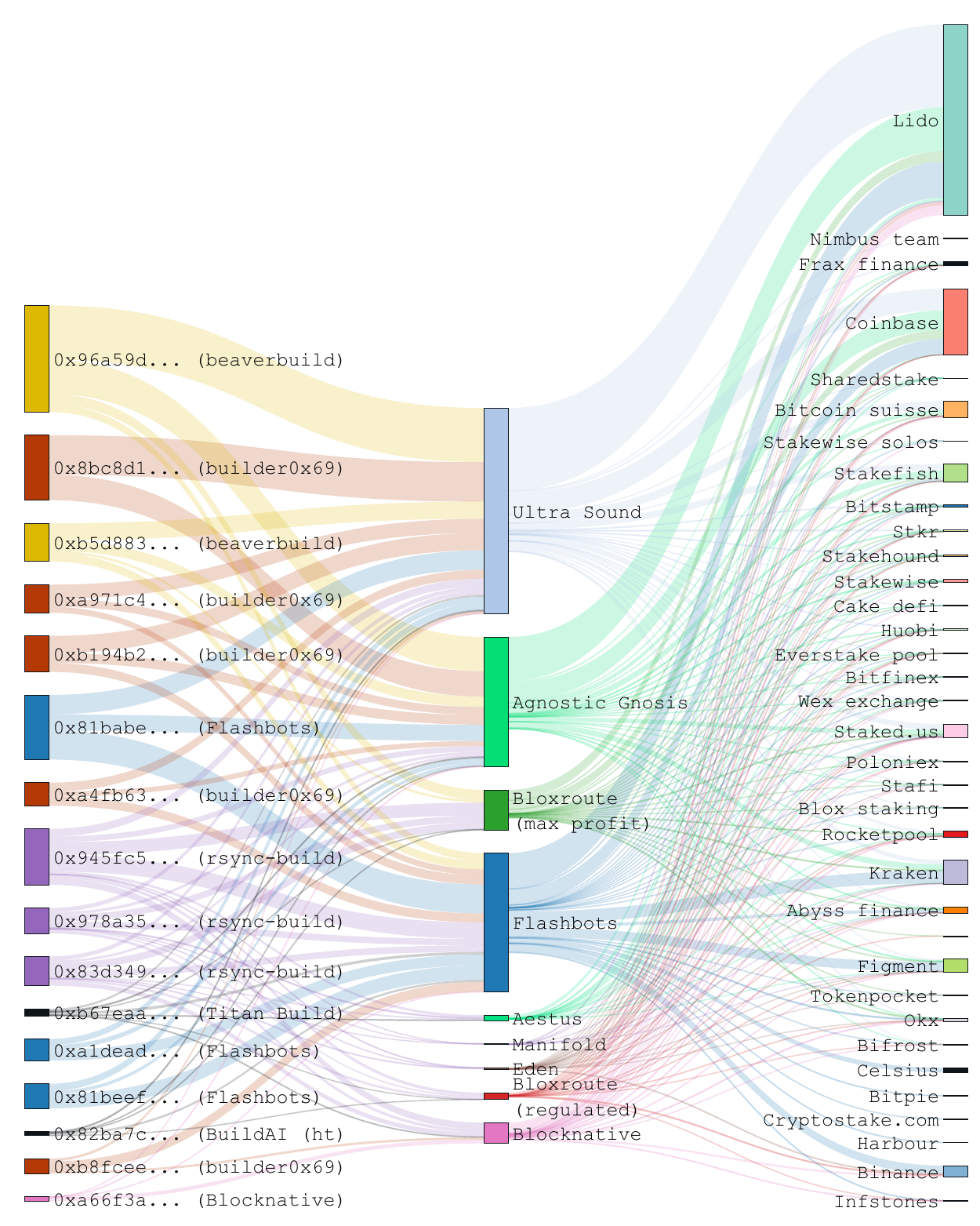}
\caption{MEV-Boost market participants, including block builders (>1500 blocks/month), relays, and proposers (left to right). The volume of the projected bands corresponds to the number of blocks for the month of April 2023.}
\label{fig:mev-boost-sankey}
\end{center}
\end{figure}

\paragraph{MEV-Boost Entity Market Distribution}
Figure~\ref{fig:mev-boost-sankey} provides a visual representation of block propagation in the MEV-Boost market, showing how blocks move from block builders (on the left) to relays and finally to Ethereum PoS proposers (on the right). For clarity, our analysis is limited to data gathered until May 24, 2023, focusing on the top 16 block builders. Each of these builders has constructed over 1500 blocks during the period from April 24 to May 24, 2023.

The current distribution of participants in the MEV-Boost market offers important insights into the interactions between different entities. For example, it reveals which relays most effectively transfer a block from a particular block builder to a specific validator. As might be anticipated, the largest proposers—such as Lido, followed by Coinbase and Kraken—amass the most blocks. Additionally, we observe that some relays, like Flashbots or Ultra Sound, maintain numerous connections. In contrast, others, such as Manifold, don't seem linked to many builders, indicating a heavy vertical integration of confident builders and relays. 

This visualization helps us better understand the dynamic interactions and connectivity patterns among various stakeholders in the MEV-Boost market. Consequently, it enables us to explore the underlying factors influencing the success and participation of entities in this rapidly evolving market.

\section{PBS Actors}\label{sec:pbs-actors}
We present our measurement results on the builders, relays, and proposers within the MEV-Boost ecosystem.

\subsection{Builders}
Builders are the participants who receive bundles of profitable MEV transactions from searchers. Their role is to assemble these transactions into profitable blocks. Once these blocks are constructed, they are sent to the relays, who pass them to the proposers.

\begin{figure}[htb]
\begin{center}
\includegraphics[width=0.95\columnwidth]{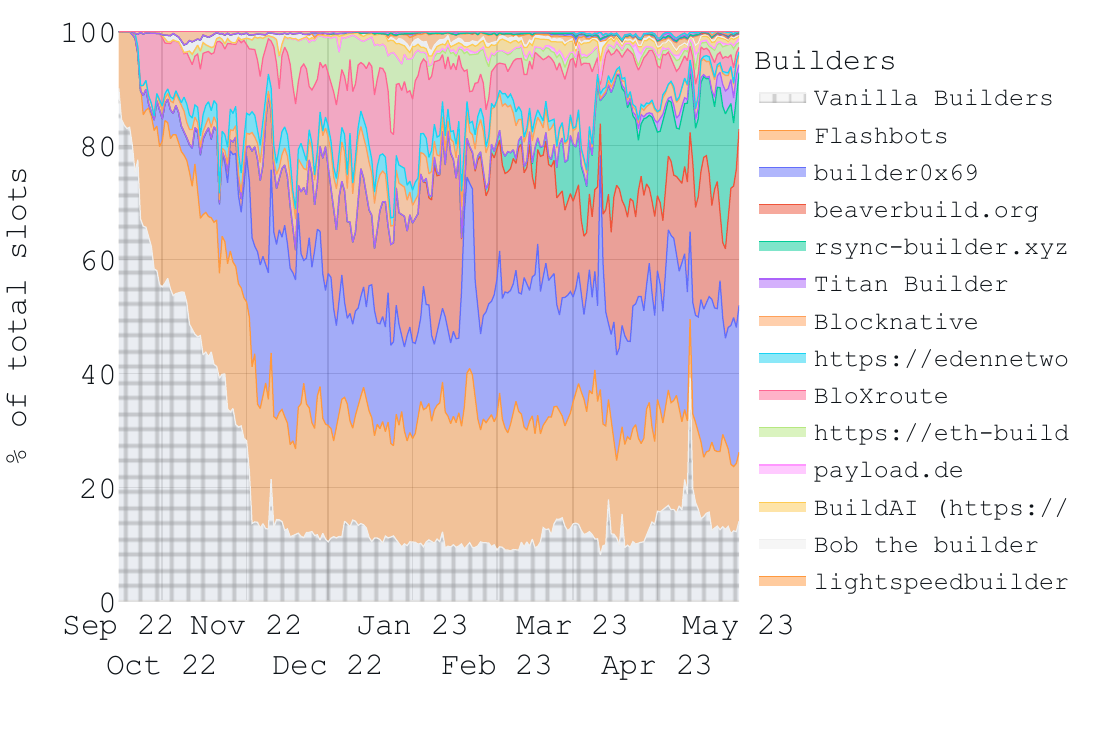}
\caption{MEV-Boost builder market share. We find that flashbots, builder0x69 and beaverbuild.org are the three biggest PBS builder.\vspace{-0.1cm}}
\label{fig:builder-market-share}
\end{center}
\end{figure}

\paragraph{Builder Distribution}
Interestingly, when we analyze the market shares of block builders, we find a consistently varied landscape throughout our observation period. No single builder holds more than 20\% of the market share at any point in time, as depicted in Figure~\ref{fig:builder-market-share}. Notable block builders include Flashbots, Beaverbuild, and Builder0x69. It's also worth mentioning that the market shares of alternative relay networks like bloXroute have fluctuated, at times capturing a significant portion only to lose it later. This observation highlights the dynamic nature of the block builder market. It is marked by the simultaneous existence of numerous players, which seems to promote healthy competition and spur innovation in the block-building ecosystem.

\paragraph{Builder MEV Payments}
We now shift our attention to the financial values associated with builder transactions. It's important to understand that builders receive bundles, which are essentially a sequence of MEV-extracting transactions, from searchers. Their main objective is to create profitable blocks for proposers, while potentially keeping a share of the rewards for themselves.

\begin{figure}[htb]
\begin{center}
\includegraphics[width=0.95\columnwidth]{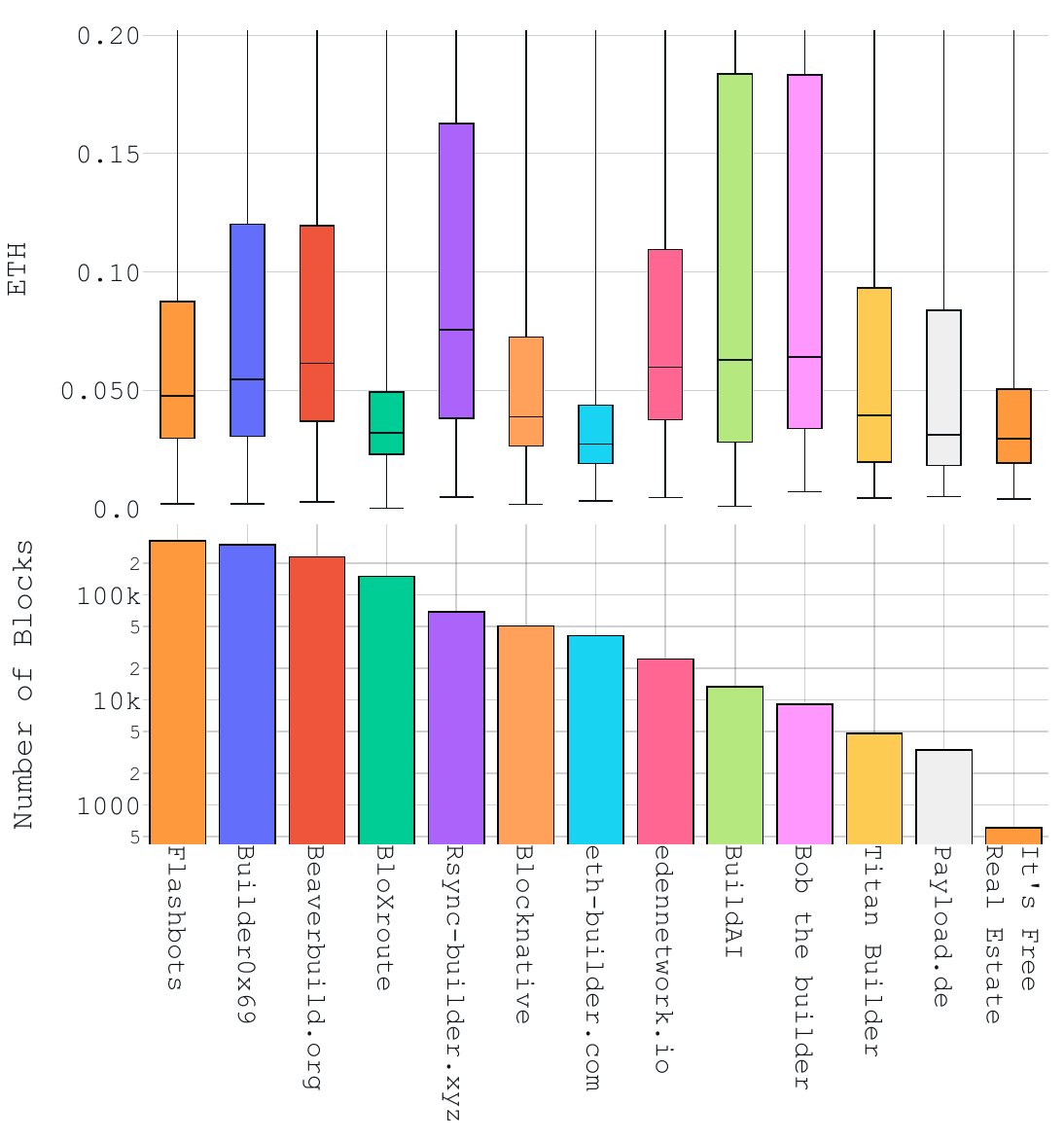}
\caption{Average MEV-Boost payments across builders. Note that the payment distribution is not very even. Sorted by size, biggest on the left, to the right.}
\label{fig:mev-boost-builderBox}
\end{center}
\end{figure}

Figure~\ref{fig:mev-boost-builderBox} illustrates the distribution of builders' ETH payments to proposers. Interestingly, there are substantial disparities among builders, and their respective market shares do not necessarily align with their payments—contrary to what one might intuitively expect. For example, Flashbots, one of the largest builders, is not the highest payer, with an average payment slightly above 0.04 ETH. Conversely, some smaller builders like BuildAI, which holds approximately 2\% market share, have a higher average payment of 0.06 ETH.

These findings emphasize the significance of effective financial incentives and factors such as ease of integration, market brand dominance, and potential latency among the actors operating in the PBS system. The dynamic interaction among these factors contributes to a more complex understanding of the relationships between builders and proposers in the Ethereum network. Builders can influence this metric by primarily building during periods of high MEV extraction.

\subsection{Relays}
\begin{figure}[htb]
\begin{center}
\includegraphics[width=0.95\columnwidth]{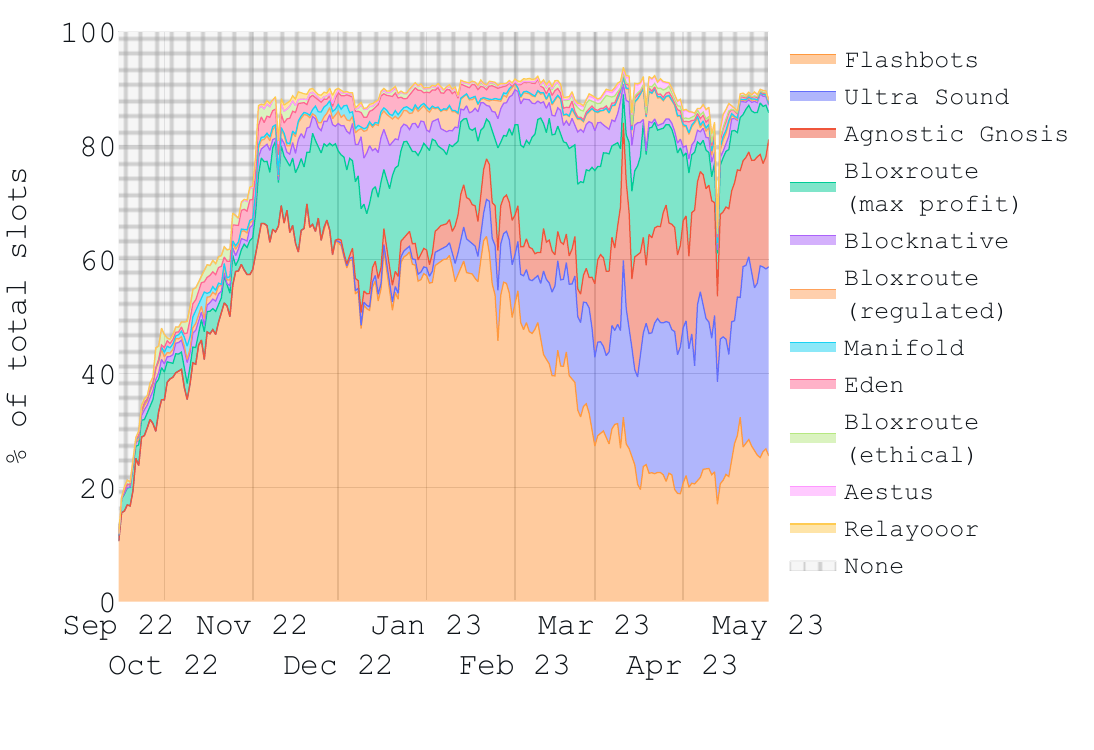}
\caption{MEV-Boost relay market share. We observe that while flashbots attracted a significant market share (60\%), likely because they were dominant before PBS, their apparent share declined recently.}
\label{fig:relayShare}
\end{center}
\end{figure}

Our focus now shifts to relays, the intermediaries connecting block builders and proposers within the PBS ecosystem. Similar to our investigation on block builders, we first examine their respective market shares. As depicted in Figure~\ref{fig:relayShare}, we make two primary observations:

\begin{enumerate}
    \item Before April 2023, the Flashbots relay expanded its market share to over 60\%, while the Flashbots builder was responsible for only 20\% of the blocks built during that time. This development is logical, as Flashbots was among the first relays to participate, granting it a significant first-mover advantage. Therefore, the first relay that proposers likely interacted with was the Flashbots relayer.
    \item However, since April 2023, the market shares of relays do not significantly deviate from those of block builders, and Flashbots's share declined to about 25\%.
\end{enumerate}

We postulate that this observation can be primarily attributed to the incentives for a block builder to operate as a block relayer, pointing to incentives for vertical integration. We identify the following three supporting factors:
\begin{itemize}
    \item By directly communicating with a vertically integrated relay, a block builder can exert more control over its market share while decreasing dependence on third-party intermediaries. This motivates the block builder to remove any unnecessary relays from the communication process.
    \item The fewer connections a block builder needs to communicate with proposers, the quicker its blocks can reach the proposer. This allows the builder more time to construct profitable blocks using transactions from searchers and to offer higher payments to proposers.
    \item The financial costs involved in operating a relay, which include server hardware, uptime, and technical personnel, are significant. As relays are not compensated in the current PBS system, it makes economic sense for block builders to integrate these costs into their operations.
\end{itemize}

By considering these factors, we can better understand the incentives driving the interplay between block builders and relays in the Ethereum network.

\paragraph{Analysis of Relay MEV Payments}
In this section, we conduct a detailed analysis of the payments that a relay transfers to proposers. There's a hypothesis that block builders are incentivized to operate relays. Our quantitative data indicates that the MEV payments proposers receive through relays seem more evenly spread among them than the individual payments originating from block builders.

\begin{figure}[htb]
\begin{center}
\includegraphics[width=0.95\columnwidth]{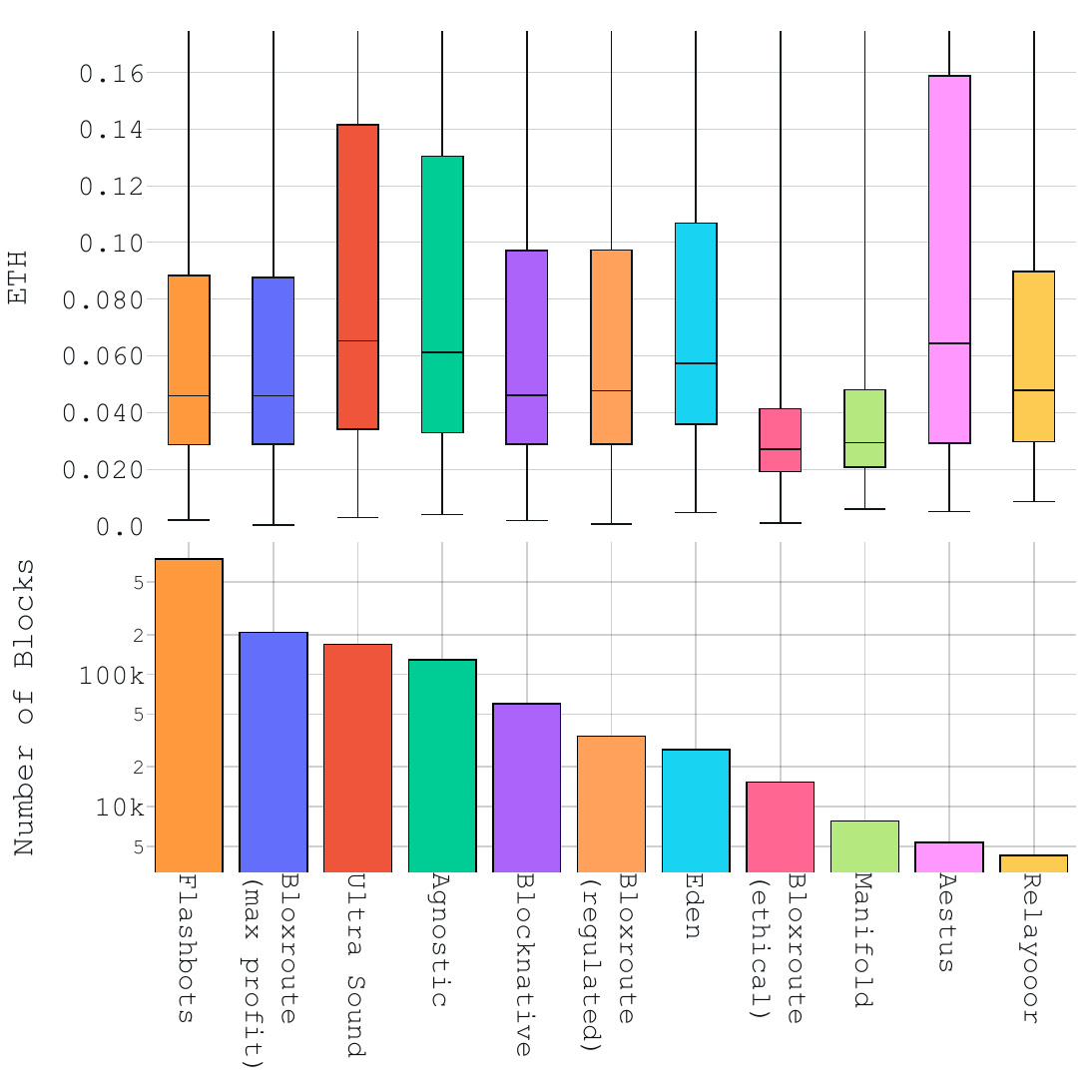}
\caption{Average MEV-Boost payments across relays. Sorted by size, biggest on the left, to the right.}
\label{fig:mevboost-relayBox}
\end{center}
\end{figure}

Figure~\ref{fig:mevboost-relayBox} shows the distribution of MEV-Boost payments made by relays to proposers. Unlike Figure~\ref{fig:mev-boost-builderBox}, which presents the payment distribution made by block builders, the MEV payments facilitated by relays exhibit a higher level of uniformity. This observation can be attributed to block builders submitting their bids to multiple relays, leading to a more balanced distribution of payments from the relays. The interconnectedness between block builders and relays is further evidenced in Figure~\ref{fig:mev-boost-sankey}, which portrays the distribution of MEV-Boost market participants and block flow.

\subsection{Proposers}

In this section, we focus on proposers, the participants responsible for choosing the blocks that will extend the Ethereum blockchain. The distribution of proposers bears similarities to the past distribution of Proof-of-Work (PoW) hashing power in Ethereum, despite the fact that the participating entities have changed and evolved over time.

Interestingly, the relative share of identified proposers belonging to prominent staking services or centralized exchanges has noticeably decreased since September 1st, 2022. This trend could be seen as a positive sign of increased decentralization within the Ethereum network. It's crucial to note that the share of ``others''—which includes smaller proposers and those not part of a staking service provider—has been increasing. On April 12th, 2023, Ethereum introduced a feature that allows proposers to withdraw their staked ETH. Following this, we observed an increase in staked capital among major proposers such as Lido and Coinbase, while others like Kraken experienced a reduction in shares. This decline may be partly due to regulatory constraints faced by some organizations. However, what is particularly noteworthy is the significant growth in the share of ``others'' proposers. Since Ethereum enabled proposers' withdrawals, the share of these more minor participants has seen an increase of nearly 10\%. This shift further suggests a potential for increased decentralization within Ethereum.

\begin{figure}[htb]
\begin{center}
\includegraphics[width=0.95\columnwidth]{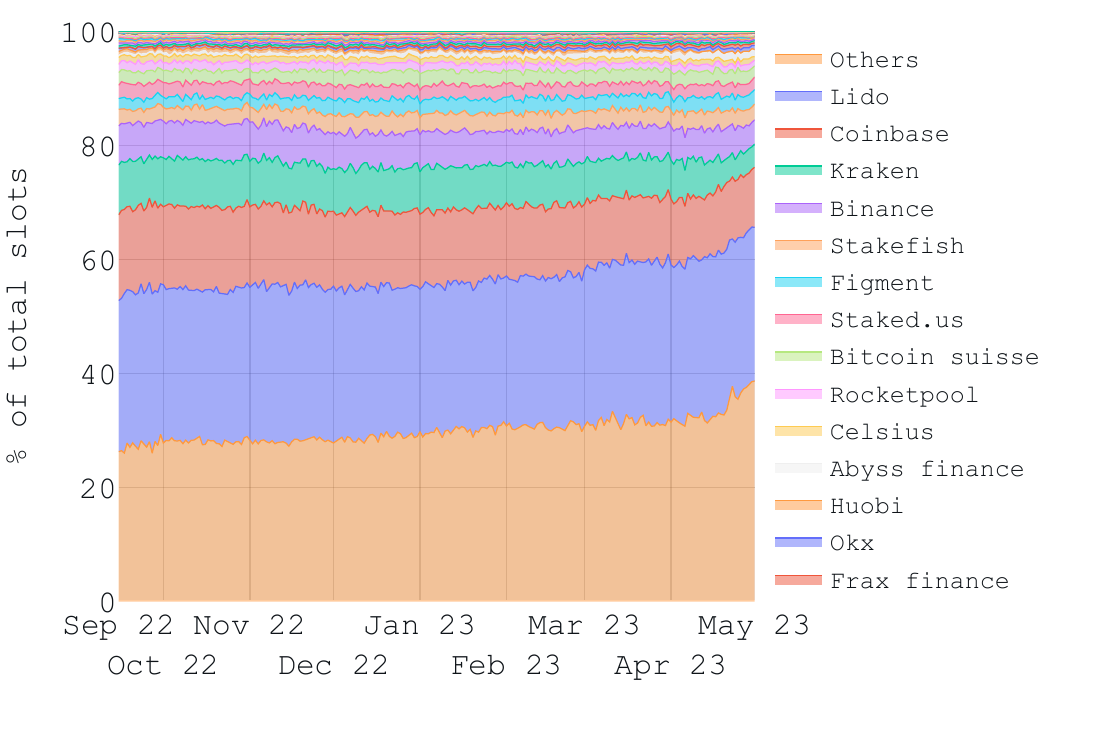}
\caption{MEV-Boost block proposer market share. We observe an increase in ``other'' proposers after Ethereum enabled the stake withdrawal in April 2023.}
\label{fig:mev-boost-validator-share}
\end{center}
\end{figure}

\paragraph{Expanding on Mined MEV Payments}
After exploring the entire process of MEV extraction, we can now focus on the complex nature of the MEV supply chain. Proposers play a crucial role in this ecosystem as they are responsible for attaching the blocks to the blockchain. Proposers chose blocks by comparing those of multiple connected relays and taking the one that yields the highest rewards for them. Interestingly, this curated selection process originates from the efficient filtering mechanism performed by relays earlier in the pipeline. The role of relays, therefore, has considerable implications for the workload of proposers, easing their burden by sorting through the candidate blocks and only presenting them with a single block. This targeted filtering allows proposers to reduce the computational complexity required, providing a streamlined method for appending suitable block candidates to the chain.

\begin{figure}[htb]
\begin{center}
\includegraphics[width=0.95\columnwidth]{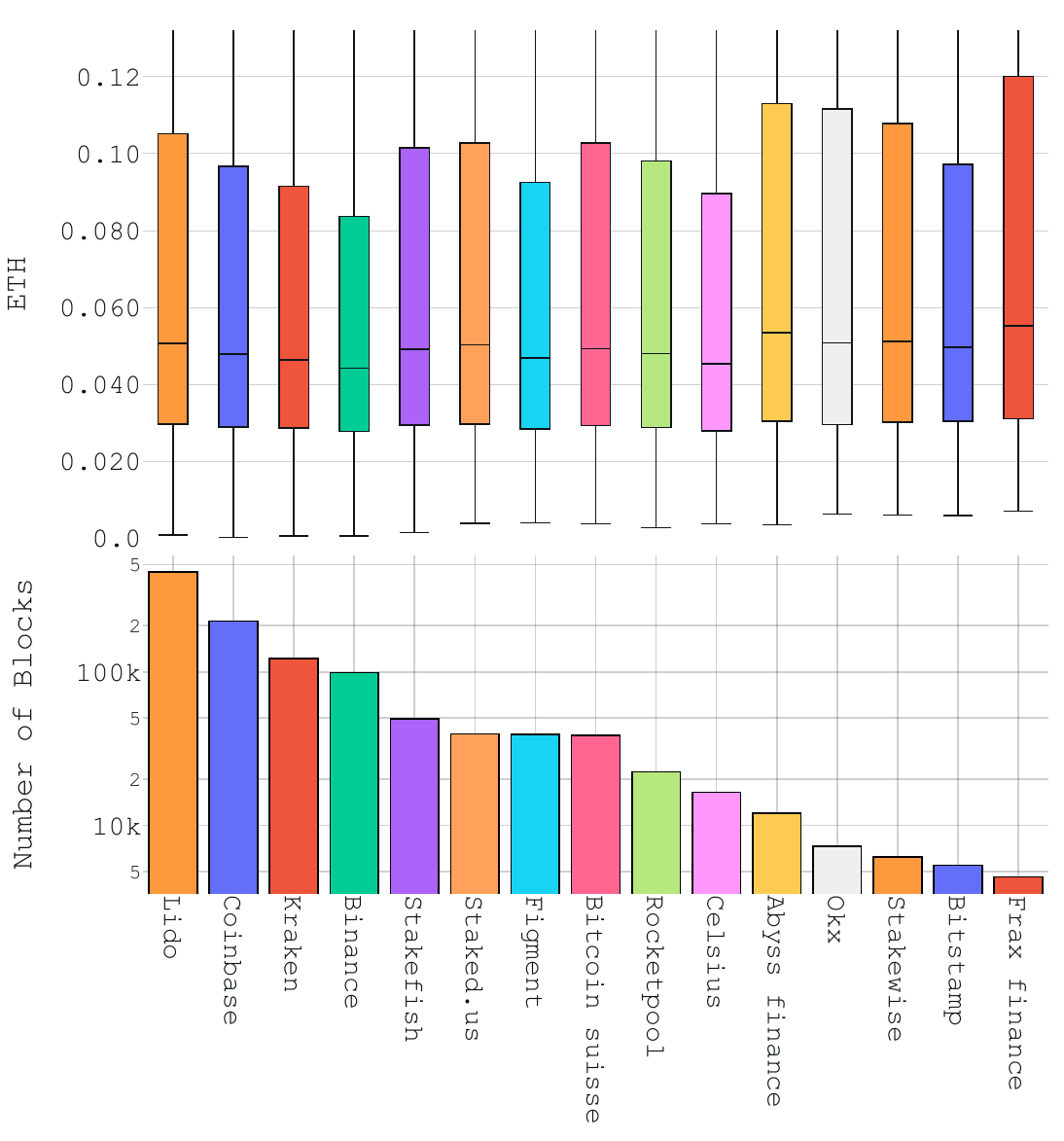}
\caption{Average MEV-Boost payments across block proposers. Sorted by size, biggest on the left, to the right.}
\label{fig:mev-boost-mined-bids}
\end{center}
\end{figure}

Insights from Figure~\ref{fig:mev-boost-mined-bids} shed light on the evolving nature of MEV payments in the blockchain ecosystem. As these payments have become more evenly distributed, the overall trend shows a leaning towards a more balanced distribution. Focusing on the largest block proposers, a very similar revenue per block can be observed.


\begin{figure}[htb]
\begin{center}
\includegraphics[width=0.95\columnwidth]{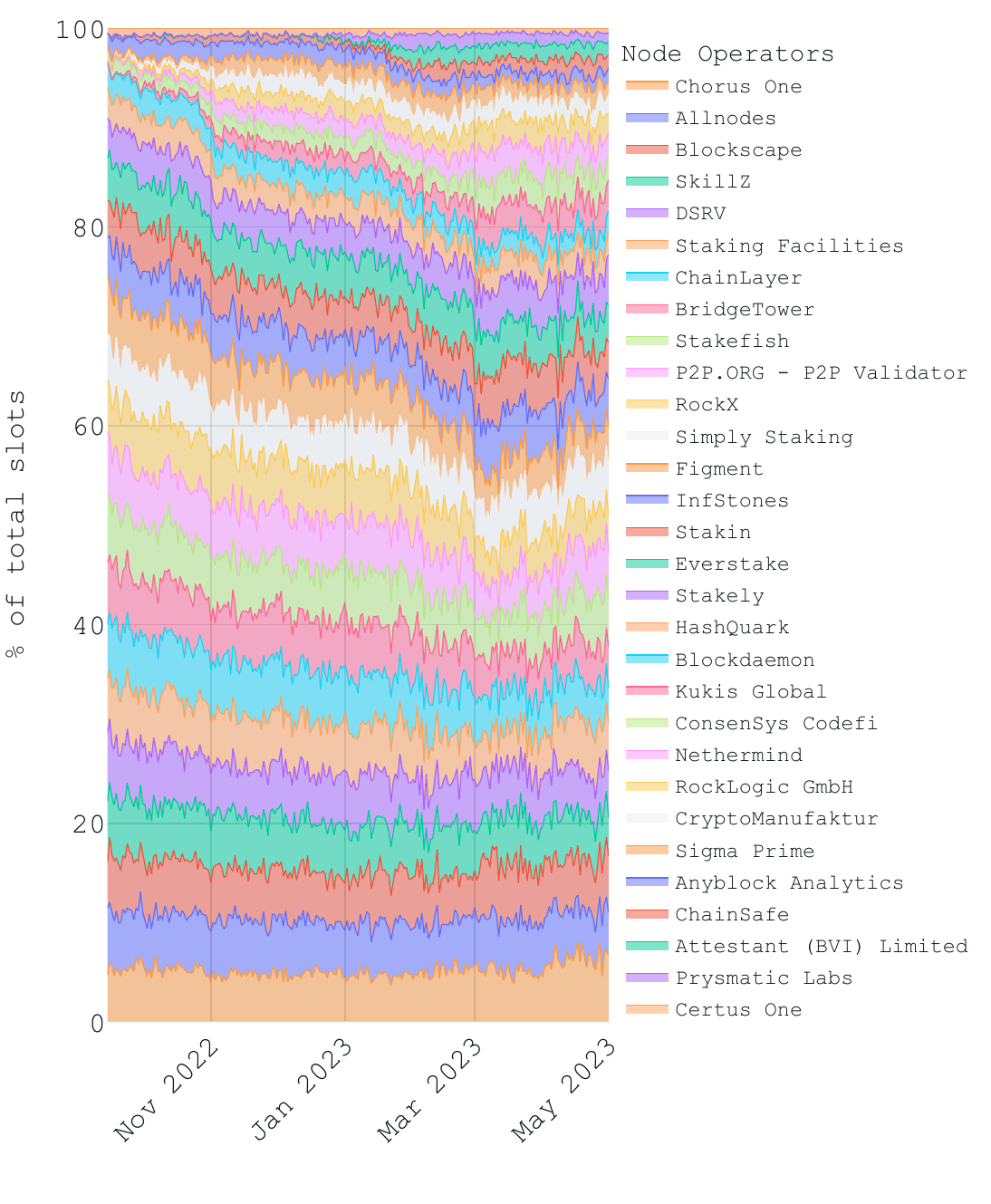}
\caption{Distribution of the 30 Lido staking operators.}
\label{fig:lido}
\end{center}
\end{figure}

\paragraph{Lido Node Operators}
Lido, a prominent liquid staking service, has emerged as the largest staking provider on Ethereum. Lido offers a staking solution that provides users with tokenized staking tokens, such as stETH. These tokens are compatible with DeFi platforms, allowing users to stake and participate in on-chain lending or trading activities simultaneously. Lido incorporates a group of node operators running block proposers. Lido Node Operators are elected by the DAO to ensure alignment with the values of the Lido DAO. Their reputation and record of past performance play a significant role in their selection. Although the number of proposers in Lido is relatively small, currently at 30, their respective stakes appear to be evenly distributed. This is visualized in Figure~\ref{fig:lido}, showing the percentage of blocks proposed by Lido's Node Operators over time.

\section{MEV-Boost Design Considerations}

An in-depth analysis and empirical review of the PBS system necessitate an objective examination of its design decisions. Ethereum's Proof of Stake system makes significant assumptions about timing constraints over an asynchronous network. Consequently, the community has observed some unexpected outcomes, which we outline below. On a broader scale, the design of MEV-Boost represents just one possible example of how to direct order flow on the network layer for MEV extraction.

\subsection{Exploring the Order Flow Design Space}
The nature of MEV inherently embodies a zero-sum game in which order flows, or user transactions subject to MEV extraction, are inevitably exposed to MEV extraction. Consequently, any proposed design is predisposed to favor certain parties over others. Achieving a democratic and fully decentralized outcome poses not only theoretical challenges but also practical hurdles. In this section, we outline the recognized boundaries of the design space.

\begin{description}
    \item[0. Price Gas Auctions] This design encompasses a single public and transparent peer-to-peer (P2P) layer where MEV auctions transpire. Participants are free to enter and exit the P2P system at will.
    \item[1. Proposer Offers MEV Endpoint] Each proposer offers an  endpoint, allowing searchers to submit bundles.
    \item[2. One-hop Intermediate Nodes: Builder/Relay] The builder and relay are managed by a single entity.
    \item[3. Two-hop Intermediate Nodes: Builder and Relay] In this design, the PBS system operates with two-hop intermediate nodes, allowing for the separation of builder and relay entities.
\end{description}

Design (0) is lauded for its full transparency, eliminating privileged access to private order flow and maintaining visibility of pending transactions. Despite this advantage, inefficiencies may arise due to counter-reactive bidding, increasing communication on the P2P network layer~\cite{daian2020flash}.

Design (1) offers simplicity as a key benefit. Precedents, such as the successful integration into PoW miners like Ethermine, support its feasibility. The potential drawback is that solo miners with limited resources and no access to sophisticated transaction ordering mechanisms may struggle to remain competitive in MEV extraction. This scenario could prompt economically rational proposers to collaborate with sophisticated centralized staking providers or compel solo validators to utilize advanced open-source/proprietary MEV extraction tools. It is worth noting that large funds are required to extract certain types of MEV, such as sandwich attacks, which may be challenging for solo miners to amass~---~a challenge which would benefit regular users. Also, cross-blockchain arbitrage or arbitrage between centralized and decentralized exchanges requires additional efforts that are potentially better handled by specialized parties.

Designs (2) and (3) aim to accommodate a PBS concept known as "weak proposer friendliness" \href{https://ethresear.ch/t/proposer-block-builder-separation-friendly-fee-market-designs/9725}{(link)}. This idea posits that proposers need not possess extensive technical skills or connections. However, this requirement necessitates the integration of technically sophisticated and well-connected intermediaries, incurring increased system latency and intermediary service fees which ultimately may reduce the optimality of MEV extraction. Also recall that the initial impetus behind blockchains was the elimination of intermediaries.

\paragraph{Conflict of Interest Considerations}
Design (0) does not present discernible conflicts of interest. However, design (1) entails potential conflicts, as proposers may double as searchers, enabling MEV theft from searchers via strategies such as the imitation game~\cite{qin2023blockchain}. It remains debatable whether this is economically rational, considering that proposers may profit more from shared MEV fees offered by external searchers. In the event of MEV theft by a proposer, searchers may have no recourse but to defer MEV to subsequent proposers. Designs (2) and (3), in which builders and relays could potentially steal MEV rewards from searchers, seem more vulnerable to conflicts of interest. Unlike proposers, builders and relays lack an ETH stake that could incur slashing consequences. While MEV theft can be detected, identifying the responsible party may prove challenging due to PBS's intricate architecture, the involvement of multiple entities, and the complex interconnectedness among builders. If a builder who commits MEV theft is identified, the searcher might redirect the MEV to alternative builders in the future.



\subsection{System Latency}
MEV-Boost added network communication and latency between searchers and proposers. This is fundamentally due to its separation of builders and relays, which were consolidated as a single entity in the initial version of Flashbots. Notably, builders can also act as relays and can even operate on the same hardware to minimize latency. Moreover, a relay may forward a block to a proposer without verifying its contents to gain a latency advantage. The more latency the system introduces, the less profitable the blocks proposers ultimately receive. Conversely, the more time a proposer waits until receiving an externally built block, the more revenue the proposer can extract.

\begin{figure*}[htb]
\begin{center}
\includegraphics[width=1.95\columnwidth]{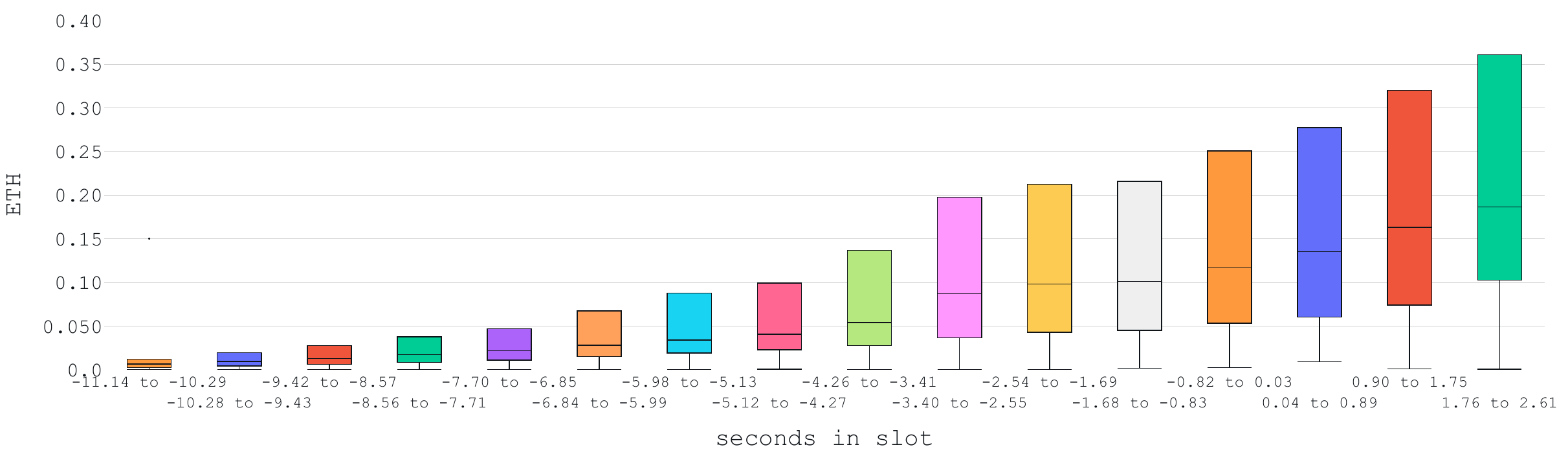}
\caption{Average MEV-Boost Payments over time arriving at the relays, relative to the start of the bid slot. Many bids arrive at the relays a few seconds before a slot starts. That is useful since the relays can then forward the most profitable bid to the proposer right when the slot starts.}
\label{fig:mev-boost-payments-time}
\end{center}
\end{figure*}

Since April 2023, we have collected the MEV-Boost bids submitted to relays along with their millisecond-precise network timestamps. Per Ethereum slot, we count an average of $1850$ bids per slot submitted to all relays. In Figure~\ref{fig:mev-boost-payments-time} we plot the distribution of the MEV-Boost bids received at the relays over the time within a slot and cluster them respectively in discrete bins since the last slot began. As expected, we observe that many bids arrive at the relays before a slot starts, which is necessary so that relays are ready to send suitable blocks to proposers once the slot starts and the proposer requests a block. The distribution of the MEV bids does not appear to be linear. More specifically, we can observe a sudden increase in the number of bids and their respective bid value about 2-3 seconds before the slot begins.

\section{Proposer Timing Discrepancies}
\begin{figure}[htb]
\begin{center}
\includegraphics[width=0.95\columnwidth]{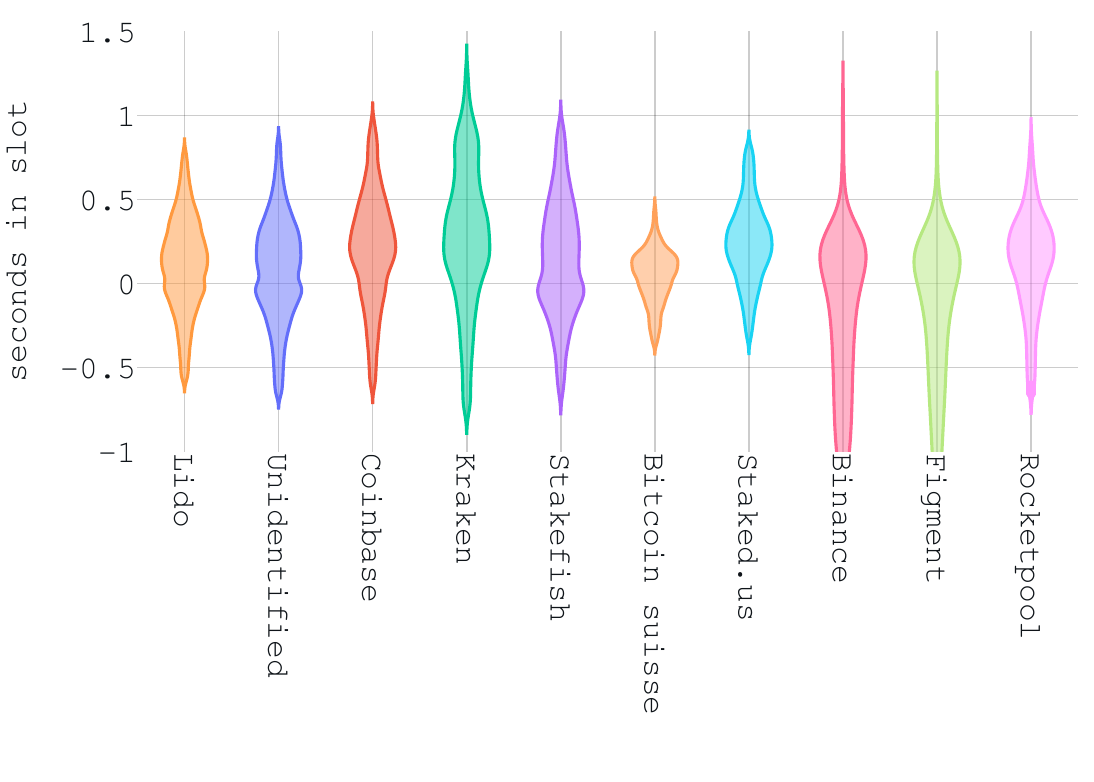}
\caption{Successful bid timing over proposers. Bitcoin Suisse appears to lead by example wrt.\ precise timing, while Kraken seems more flexible, considering bids that are submitted relatively later to relays.}
\label{fig:mev-boost-timing-validators}
\end{center}
\end{figure}

\begin{figure}[htb]
\begin{center}
\includegraphics[width=0.95\columnwidth]{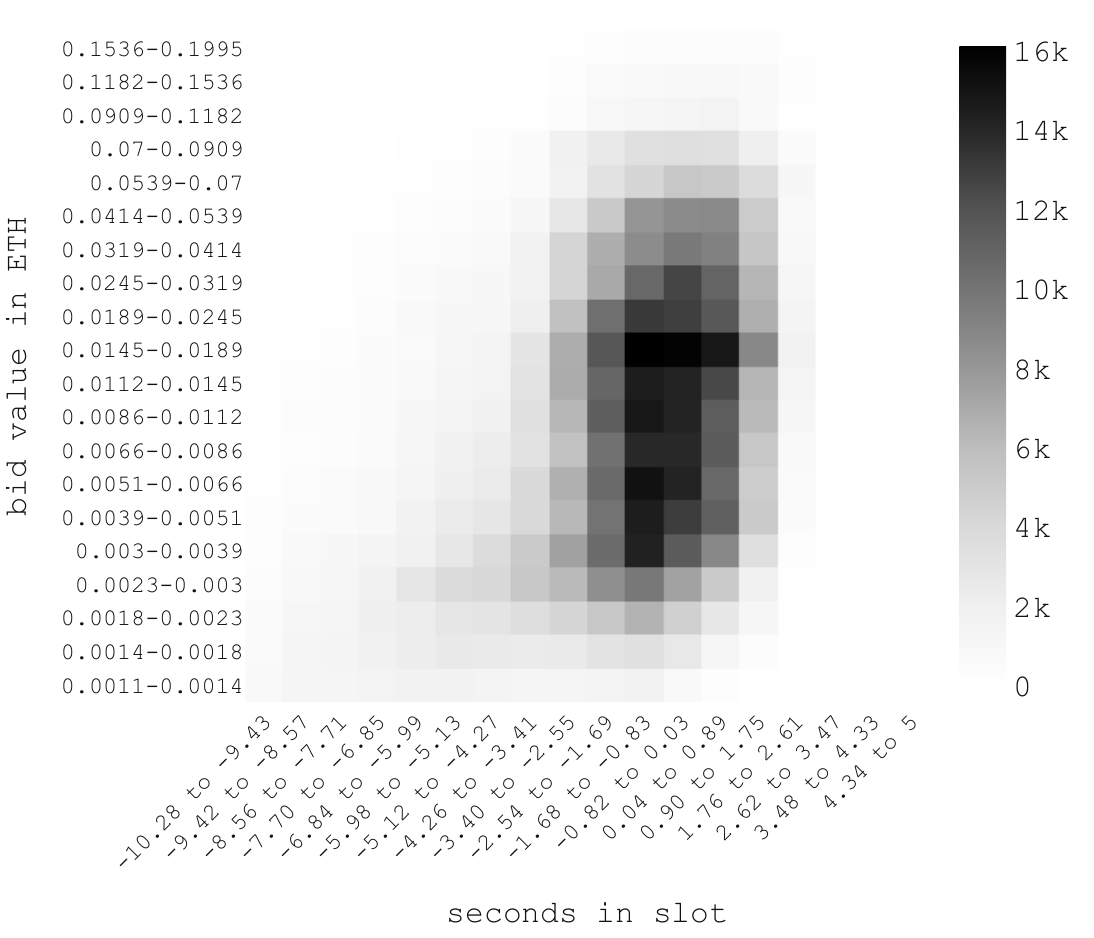}
\caption{MEV-Boost bid distribution across value, time in the slot, and frequency. The value of a bid slowly grows to anticipate a slot start, and then increases around one second before the start of the slot.}
\label{fig:mev-boost-bid-dist}
\end{center}
\end{figure}

\begin{figure}[htb]
\begin{center}
\includegraphics[width=0.95\columnwidth]{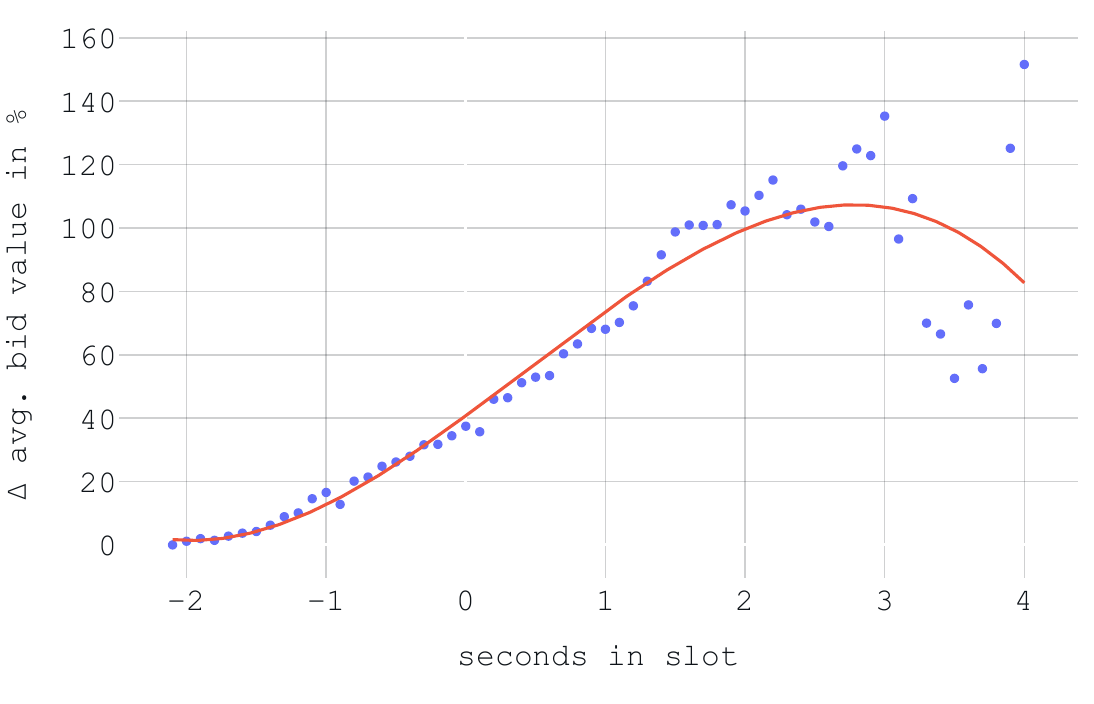}
\caption{MEV-Boost bid distribution over time and relative to the average bid value at slot second -2. At a slot age of 2.78 seconds, we find the maximum, in which the average bid value is 107 \% higher than the average bid at second -2.}
\label{fig:mev-boost-bid-regression}
\end{center}
\end{figure}

In this section, we aim to provide a comprehensive investigation into the disparities in the timing of block bids submitted by builders to the relay. We focus our quantitative analysis on successful block bids---those that eventually contained a block hash of a block that made it on-chain. In particular, we will examine the significance of the arrival time of block bids and explore its correlation with the bid value (cf.\ Figure~\ref{fig:mev-boost-timing-validators}, ~\ref{fig:mev-boost-bid-dist}, and ~\ref{fig:mev-boost-bid-regression}).

Figure~\ref{fig:mev-boost-timing-validators} illustrates variations in the timing distribution of bids. For instance, Bitcoin Suisse stands out as a very consistent proposer, predominantly using bids that arrive at the relay within the first 0.5 seconds after a slot begins. In contrast, Kraken's timing pattern seems more lenient, as the bids for proposed blocks often reach the relay over 1 second after the slot starts. These differences suggest diverse strategies used by different proposers. 

Alongside Figure~\ref{fig:mev-boost-timing-validators}, Figure~\ref{fig:mev-boost-bid-dist} displays bid values together with their corresponding timestamps upon reaching the relay. Key insights from this figure uncover a noticeable relationship between a bid's arrival time and its value within a slot.

Delving deeper into the correlation between bid arrival time and bid value, we extract Figure~\ref{fig:mev-boost-bid-regression}, which reflects the changes in the average bid value relative to the average bid value at slot second -2, our selected baseline. As indicated in both Figures~\ref{fig:mev-boost-payments-time} and \ref{fig:mev-boost-bid-dist}, slot second -2 comprises a significant number of incoming bids. A polynomial regression captures the relationship between bid time and value, resulting in an empirical MEV-time law (Equation~\ref{eq:max-bid}) to guide proposers in optimizing wait time to maximize MEV rewards.

\begin{equation}\label{eq:max-bid}
-1.99x^3 + 2.44x^2 + 32.5x + 40.77
\end{equation}

Our findings suggest that the optimal slot time for maximizing bid value is 2.78 seconds, resulting in a relative bid value that is 107\% higher than the average at second -2. However, trying to achieve this optimal waiting time to capitalize on MEV rewards can pose risks to the proposer's block. Specifically, a longer waiting period might lead to an insufficient number of attestations, which can negatively impact the network's confidence in the block's validity. Even more critically, if the waiting time is too long, the block could get reorganized, or ``re-orged'', by the subsequent proposer. This could result in the original block being discarded and replaced with a new block, causing the original proposer to lose out on potential rewards. Therefore, while there are potential financial benefits to maximizing waiting time, there is also a balancing act that proposers must manage to ensure their blocks are both profitable and stable. This dynamic interplay between profitability and stability is an essential aspect of the MEV landscape on Ethereum.

In conclusion, while our analysis offers insights into the optimal waiting time for proposers to maximize MEV rewards, it's essential to acknowledge that we have not analyzed the attestation data necessary to determine the perfect balance between optimizing bid value and ensuring block validity. Nevertheless, the information provided illustrates a noticeable relationship between bid value and timing, which can inform proposers' decision-making processes. Additionally, the optimal waiting time will also depend on the connectivity of the proposer and the relay within the network to ensure that blocks are propagated quickly and efficiently to collect a sufficient number of attestations until the attestation deadline at the second 4 in the slot.

\section{Related Works}\label{sec:relatedworks}
The study of decentralized ledgers, particularly Bitcoin and Ethereum, has been a significant area of research over the past decade. Several works have explored the peer-to-peer networks these systems rely on for transaction data propagation, examining their efficiency and security~\cite{gervais2015tampering}.

Overall, the study of Miner Extractable Value has gained considerable attention in recent years, particularly in the context of Ethereum's transition from Proof of Work to Proof of Stake. Daian et al.\ provided an early analysis of MEV, examining its origins, implications, and possible future developments~\cite{daian2020flash}. Qin et al.\ expanded on this analysis, detailing the various forms of MEV and their impact on blockchain security and fairness~\cite{qin2022quantifying}. The evolution of these systems, including the introduction of ports for direct transaction data submission by miners and the emergence of centralized intermediaries like Flashbots, has also been a topic of interest. These developments have significantly transformed the information dissemination layer of decentralized ledgers. Our paper complements these studies by investigating the effects of the Proposer Builder Separation mechanism on the Ethereum block construction market.

Roughgarden~\cite{roughgarden2021transaction} explores the design of blockchain transaction fee mechanisms, focusing on incentive compatibility in the context of blockchains like Bitcoin and Ethereum. The authors introduce two new incentive-compatibility forms: MMIC, which protects against profit-maximizing miner deviations, and OCA-proofness, which protects against off-chain collusion between miners and users. The study is relevant to MEV and Ethereum's recent major change in its transaction fee mechanism (EIP-1559), which replaced the first-price auction with a system that includes variable-size blocks, history-dependent reserve prices, and transaction fee burning.

This paper focuses on the Ethereum block construction market and the impact of the Proposer Builder Separation mechanism following the introduction of PoS and PBS in September 2022. We analyze the market shares of builders and relays, temporal changes, and financial dynamics.

The analysis of blockchain crises, such as the FTX collapse and the USDC stablecoin de-peg, was the subject of previous preliminary research. For instance, possible DeFi crisis and its implications for DeFi protocols were analyzed shallowly~\cite{gudgeon2020decentralized}. We are the first to examine the quantitative relationship between crisis events and spikes in MEV payments.

Finally, the design of blockchain systems has been investigated by various scholars. Gervais et al.\ explored the trade-offs between security and performance in blockchain systems \cite{gervais2016security}. Our paper contributes to this body of work by raising questions about the optimality of the PBS design and its potential regulatory implications, e.g., around the privileges of private order flow that MEV-Boost nurtures.

\section{Conclusion}\label{sec:conclusion}
The Ethereum community deserves recognition for its proactive approach and implementation of considerable changes, such as the Proposer Builder Separation, which foster proposer decentralization. Through PBS, the system removes the authority of proposers to order transactions and, thus, MEV extraction. However, an interesting paradox arises from this model, where proposers commit to a block header before having its content. This process implies a level of trust placed on block builders and relayers, which seems to contradict the decentralized ethos of the system, considering these actors could potentially manipulate MEV rewards through generalized front-running strategies.

The justification behind this implicit trust isn't immediately apparent, particularly given the potential for block builders and relayers to operate their own searchers to extract MEV. Moreover, while the PBS model has seen widespread acceptance in the Ethereum ecosystem, largely due to the backing from key stakeholders, it's worth noting that this model hasn't been universally adopted across other blockchain networks. Alternate MEV auction designs may be more effective, e.g., where the proposer directly opens an endpoint for receiving blocks, eliminating the need for trusted intermediaries.

In summary, it's crucial to have a comprehensive understanding of the intricate mechanisms and interactions governing MEV payments for the evolution and enhancement of the Ethereum blockchain. 

\bibliographystyle{ACM-Reference-Format}
\bibliography{references}

\end{document}